\documentclass[longauth]{aaEC}  
\usepackage{graphicx}
\usepackage{natbib}
\usepackage{twoopt}
\usepackage{txfonts}
\usepackage{xcolor}
\usepackage{placeins}

\newcommand{\ev}[1]{\langle #1\rangle}
\newcommand{\hinvmpc}{h^{-1}{\rm Mpc}}
\newcommand{\Minv}{M^{-1}}
\newcommand{\Msun}{M_{\odot}}
\newcommand{\N}[1]{N_{\rm #1}}
\newcommand{\Om}{\Omega_{\rm m}}
\newcommand{\Pm}{P_{\rm m}(k)}
\newcommand{\thetab}{\vec{\Theta}}
\newcommand{\datab}{\vec{d}}
\newcommand{\modelb}{\vec{m}(\thetab)}
\newcommand{\covb}{\mathbf{C}(\thetab)}
\newcommand{\invcovb}{\mathbf{C}^{-1}(\thetab)}

\usepackage[pdfencoding=auto,psdextra]{hyperref}
\hypersetup{
    colorlinks=true,
    linkcolor=blue,
    filecolor=magenta,      
    urlcolor=blue,
    citecolor=blue
}

\usepackage{euclid}
\usepackage[switch, modulo]{lineno}
\makeatletter
\renewcommand*\aa@pageof{, page \thepage{} of \pageref*{LastPage}}
\makeatother

\begin{document}

   \title{\Euclid\/: The linear-construction covariance and cosmology\thanks{This paper is published on
     behalf of the Euclid Consortium}}

   \subtitle{}

%%%% Version Friday 6th of March 2026 08:02:04 AM UT												
%%%% Please do not edit the author list -- contact ECEB Bureau for changes
\newcommand{\orcid}[1]{} %% if already defined in aa.cls: comment, or use renewcommand			   
\author{V.~Lindholm\orcid{0000-0003-2317-5471}\thanks{\email{valtteri.lindholm@helsinki.fi}}\inst{\ref{aff1},\ref{aff2}}
\and E.~Sihvola\orcid{0000-0003-1804-7715}\inst{\ref{aff3}}
\and J.~Valiviita\orcid{0000-0001-6225-3693}\inst{\ref{aff1},\ref{aff2}}
\and A.~Fumagalli\orcid{0009-0004-0300-2535}\inst{\ref{aff4}}
\and B.~Altieri\orcid{0000-0003-3936-0284}\inst{\ref{aff5}}
\and S.~Andreon\orcid{0000-0002-2041-8784}\inst{\ref{aff6}}
\and N.~Auricchio\orcid{0000-0003-4444-8651}\inst{\ref{aff7}}
\and C.~Baccigalupi\orcid{0000-0002-8211-1630}\inst{\ref{aff8},\ref{aff4},\ref{aff9},\ref{aff10}}
\and M.~Baldi\orcid{0000-0003-4145-1943}\inst{\ref{aff11},\ref{aff7},\ref{aff12}}
\and S.~Bardelli\orcid{0000-0002-8900-0298}\inst{\ref{aff7}}
\and P.~Battaglia\orcid{0000-0002-7337-5909}\inst{\ref{aff7}}
\and A.~Biviano\orcid{0000-0002-0857-0732}\inst{\ref{aff4},\ref{aff8}}
\and E.~Branchini\orcid{0000-0002-0808-6908}\inst{\ref{aff13},\ref{aff14},\ref{aff6}}
\and M.~Brescia\orcid{0000-0001-9506-5680}\inst{\ref{aff15},\ref{aff16}}
\and S.~Camera\orcid{0000-0003-3399-3574}\inst{\ref{aff17},\ref{aff18},\ref{aff19}}
\and V.~Capobianco\orcid{0000-0002-3309-7692}\inst{\ref{aff19}}
\and C.~Carbone\orcid{0000-0003-0125-3563}\inst{\ref{aff20}}
\and V.~F.~Cardone\inst{\ref{aff21},\ref{aff22}}
\and J.~Carretero\orcid{0000-0002-3130-0204}\inst{\ref{aff23},\ref{aff24}}
\and S.~Casas\orcid{0000-0002-4751-5138}\inst{\ref{aff25},\ref{aff26}}
\and M.~Castellano\orcid{0000-0001-9875-8263}\inst{\ref{aff21}}
\and G.~Castignani\orcid{0000-0001-6831-0687}\inst{\ref{aff7}}
\and S.~Cavuoti\orcid{0000-0002-3787-4196}\inst{\ref{aff16},\ref{aff27}}
\and K.~C.~Chambers\orcid{0000-0001-6965-7789}\inst{\ref{aff28}}
\and A.~Cimatti\inst{\ref{aff29}}
\and C.~Colodro-Conde\inst{\ref{aff30}}
\and G.~Congedo\orcid{0000-0003-2508-0046}\inst{\ref{aff31}}
\and L.~Conversi\orcid{0000-0002-6710-8476}\inst{\ref{aff32},\ref{aff5}}
\and Y.~Copin\orcid{0000-0002-5317-7518}\inst{\ref{aff33}}
\and F.~Courbin\orcid{0000-0003-0758-6510}\inst{\ref{aff34},\ref{aff35},\ref{aff36}}
\and H.~M.~Courtois\orcid{0000-0003-0509-1776}\inst{\ref{aff37}}
\and A.~Da~Silva\orcid{0000-0002-6385-1609}\inst{\ref{aff38},\ref{aff39}}
\and H.~Degaudenzi\orcid{0000-0002-5887-6799}\inst{\ref{aff40}}
\and G.~De~Lucia\orcid{0000-0002-6220-9104}\inst{\ref{aff4}}
\and H.~Dole\orcid{0000-0002-9767-3839}\inst{\ref{aff41}}
\and F.~Dubath\orcid{0000-0002-6533-2810}\inst{\ref{aff40}}
\and X.~Dupac\inst{\ref{aff5}}
\and S.~Dusini\orcid{0000-0002-1128-0664}\inst{\ref{aff42}}
\and S.~Escoffier\orcid{0000-0002-2847-7498}\inst{\ref{aff43}}
\and M.~Farina\orcid{0000-0002-3089-7846}\inst{\ref{aff44}}
\and R.~Farinelli\inst{\ref{aff7}}
\and S.~Ferriol\inst{\ref{aff33}}
\and F.~Finelli\orcid{0000-0002-6694-3269}\inst{\ref{aff7},\ref{aff45}}
\and P.~Fosalba\orcid{0000-0002-1510-5214}\inst{\ref{aff46},\ref{aff47}}
\and S.~Fotopoulou\orcid{0000-0002-9686-254X}\inst{\ref{aff48}}
\and M.~Frailis\orcid{0000-0002-7400-2135}\inst{\ref{aff4}}
\and E.~Franceschi\orcid{0000-0002-0585-6591}\inst{\ref{aff7}}
\and M.~Fumana\orcid{0000-0001-6787-5950}\inst{\ref{aff20}}
\and S.~Galeotta\orcid{0000-0002-3748-5115}\inst{\ref{aff4}}
\and K.~George\orcid{0000-0002-1734-8455}\inst{\ref{aff49}}
\and B.~Gillis\orcid{0000-0002-4478-1270}\inst{\ref{aff31}}
\and C.~Giocoli\orcid{0000-0002-9590-7961}\inst{\ref{aff7},\ref{aff12}}
\and J.~Gracia-Carpio\orcid{0000-0003-4689-3134}\inst{\ref{aff50}}
\and A.~Grazian\orcid{0000-0002-5688-0663}\inst{\ref{aff51}}
\and F.~Grupp\inst{\ref{aff50},\ref{aff52}}
\and S.~V.~H.~Haugan\orcid{0000-0001-9648-7260}\inst{\ref{aff53}}
\and W.~Holmes\inst{\ref{aff54}}
\and F.~Hormuth\inst{\ref{aff55}}
\and A.~Hornstrup\orcid{0000-0002-3363-0936}\inst{\ref{aff56},\ref{aff57}}
\and K.~Jahnke\orcid{0000-0003-3804-2137}\inst{\ref{aff58}}
\and M.~Jhabvala\inst{\ref{aff59}}
\and S.~Kermiche\orcid{0000-0002-0302-5735}\inst{\ref{aff43}}
\and A.~Kiessling\orcid{0000-0002-2590-1273}\inst{\ref{aff54}}
\and B.~Kubik\orcid{0009-0006-5823-4880}\inst{\ref{aff33}}
\and M.~Kunz\orcid{0000-0002-3052-7394}\inst{\ref{aff60}}
\and H.~Kurki-Suonio\orcid{0000-0002-4618-3063}\inst{\ref{aff1},\ref{aff2}}
\and A.~M.~C.~Le~Brun\orcid{0000-0002-0936-4594}\inst{\ref{aff61}}
\and S.~Ligori\orcid{0000-0003-4172-4606}\inst{\ref{aff19}}
\and P.~B.~Lilje\orcid{0000-0003-4324-7794}\inst{\ref{aff53}}
\and I.~Lloro\orcid{0000-0001-5966-1434}\inst{\ref{aff62}}
\and G.~Mainetti\orcid{0000-0003-2384-2377}\inst{\ref{aff63}}
\and E.~Maiorano\orcid{0000-0003-2593-4355}\inst{\ref{aff7}}
\and O.~Mansutti\orcid{0000-0001-5758-4658}\inst{\ref{aff4}}
\and S.~Marcin\inst{\ref{aff64}}
\and O.~Marggraf\orcid{0000-0001-7242-3852}\inst{\ref{aff65}}
\and M.~Martinelli\orcid{0000-0002-6943-7732}\inst{\ref{aff21},\ref{aff22}}
\and N.~Martinet\orcid{0000-0003-2786-7790}\inst{\ref{aff66}}
\and F.~Marulli\orcid{0000-0002-8850-0303}\inst{\ref{aff67},\ref{aff7},\ref{aff12}}
\and R.~J.~Massey\orcid{0000-0002-6085-3780}\inst{\ref{aff68}}
\and E.~Medinaceli\orcid{0000-0002-4040-7783}\inst{\ref{aff7}}
\and S.~Mei\orcid{0000-0002-2849-559X}\inst{\ref{aff69},\ref{aff70}}
\and M.~Melchior\inst{\ref{aff71}}
\and M.~Meneghetti\orcid{0000-0003-1225-7084}\inst{\ref{aff7},\ref{aff12}}
\and E.~Merlin\orcid{0000-0001-6870-8900}\inst{\ref{aff21}}
\and G.~Meylan\inst{\ref{aff72}}
\and A.~Mora\orcid{0000-0002-1922-8529}\inst{\ref{aff73}}
\and M.~Moresco\orcid{0000-0002-7616-7136}\inst{\ref{aff67},\ref{aff7}}
\and L.~Moscardini\orcid{0000-0002-3473-6716}\inst{\ref{aff67},\ref{aff7},\ref{aff12}}
\and R.~Nakajima\orcid{0009-0009-1213-7040}\inst{\ref{aff65}}
\and C.~Neissner\orcid{0000-0001-8524-4968}\inst{\ref{aff74},\ref{aff24}}
\and S.-M.~Niemi\orcid{0009-0005-0247-0086}\inst{\ref{aff75}}
\and C.~Padilla\orcid{0000-0001-7951-0166}\inst{\ref{aff74}}
\and S.~Paltani\orcid{0000-0002-8108-9179}\inst{\ref{aff40}}
\and F.~Pasian\orcid{0000-0002-4869-3227}\inst{\ref{aff4}}
\and K.~Pedersen\inst{\ref{aff76}}
\and V.~Pettorino\orcid{0000-0002-4203-9320}\inst{\ref{aff75}}
\and S.~Pires\orcid{0000-0002-0249-2104}\inst{\ref{aff77}}
\and G.~Polenta\orcid{0000-0003-4067-9196}\inst{\ref{aff78}}
\and M.~Poncet\inst{\ref{aff79}}
\and L.~A.~Popa\inst{\ref{aff80}}
\and F.~Raison\orcid{0000-0002-7819-6918}\inst{\ref{aff50}}
\and A.~Renzi\orcid{0000-0001-9856-1970}\inst{\ref{aff81},\ref{aff42},\ref{aff7}}
\and J.~Rhodes\orcid{0000-0002-4485-8549}\inst{\ref{aff54}}
\and G.~Riccio\inst{\ref{aff16}}
\and E.~Romelli\orcid{0000-0003-3069-9222}\inst{\ref{aff4}}
\and M.~Roncarelli\orcid{0000-0001-9587-7822}\inst{\ref{aff7}}
\and C.~Rosset\orcid{0000-0003-0286-2192}\inst{\ref{aff69}}
\and R.~Saglia\orcid{0000-0003-0378-7032}\inst{\ref{aff52},\ref{aff50}}
\and Z.~Sakr\orcid{0000-0002-4823-3757}\inst{\ref{aff82},\ref{aff83},\ref{aff84}}
\and A.~G.~S\'anchez\orcid{0000-0003-1198-831X}\inst{\ref{aff50}}
\and D.~Sapone\orcid{0000-0001-7089-4503}\inst{\ref{aff85}}
\and P.~Schneider\orcid{0000-0001-8561-2679}\inst{\ref{aff65}}
\and T.~Schrabback\orcid{0000-0002-6987-7834}\inst{\ref{aff86}}
\and A.~Secroun\orcid{0000-0003-0505-3710}\inst{\ref{aff43}}
\and G.~Seidel\orcid{0000-0003-2907-353X}\inst{\ref{aff58}}
\and P.~Simon\inst{\ref{aff65}}
\and C.~Sirignano\orcid{0000-0002-0995-7146}\inst{\ref{aff81},\ref{aff42}}
\and G.~Sirri\orcid{0000-0003-2626-2853}\inst{\ref{aff12}}
\and L.~Stanco\orcid{0000-0002-9706-5104}\inst{\ref{aff42}}
\and P.~Tallada-Cresp\'{i}\orcid{0000-0002-1336-8328}\inst{\ref{aff23},\ref{aff24}}
\and A.~N.~Taylor\inst{\ref{aff31}}
\and I.~Tereno\orcid{0000-0002-4537-6218}\inst{\ref{aff38},\ref{aff87}}
\and S.~Toft\orcid{0000-0003-3631-7176}\inst{\ref{aff88},\ref{aff89}}
\and R.~Toledo-Moreo\orcid{0000-0002-2997-4859}\inst{\ref{aff90}}
\and F.~Torradeflot\orcid{0000-0003-1160-1517}\inst{\ref{aff24},\ref{aff23}}
\and I.~Tutusaus\orcid{0000-0002-3199-0399}\inst{\ref{aff47},\ref{aff46},\ref{aff83}}
\and T.~Vassallo\orcid{0000-0001-6512-6358}\inst{\ref{aff4},\ref{aff49}}
\and G.~Verdoes~Kleijn\orcid{0000-0001-5803-2580}\inst{\ref{aff91}}
\and Y.~Wang\orcid{0000-0002-4749-2984}\inst{\ref{aff92}}
\and J.~Weller\orcid{0000-0002-8282-2010}\inst{\ref{aff52},\ref{aff50}}
\and G.~Zamorani\orcid{0000-0002-2318-301X}\inst{\ref{aff7}}
\and E.~Zucca\orcid{0000-0002-5845-8132}\inst{\ref{aff7}}
\and T.~Castro\orcid{0000-0002-6292-3228}\inst{\ref{aff4},\ref{aff9},\ref{aff8},\ref{aff93}}
\and J.~Mart\'{i}n-Fleitas\orcid{0000-0002-8594-569X}\inst{\ref{aff94}}
\and P.~Monaco\orcid{0000-0003-2083-7564}\inst{\ref{aff95},\ref{aff4},\ref{aff9},\ref{aff8}}
\and A.~Pezzotta\orcid{0000-0003-0726-2268}\inst{\ref{aff6}}
\and V.~Scottez\orcid{0009-0008-3864-940X}\inst{\ref{aff96},\ref{aff97}}
\and M.~Sereno\orcid{0000-0003-0302-0325}\inst{\ref{aff7},\ref{aff12}}
\and M.~Viel\orcid{0000-0002-2642-5707}\inst{\ref{aff8},\ref{aff4},\ref{aff10},\ref{aff9},\ref{aff93}}
\and D.~Sciotti\orcid{0009-0008-4519-2620}\inst{\ref{aff21},\ref{aff22}}}
										   
%%%% please do not edit the affiliation list -- contact ECEB Bureau for changes
\institute{Department of Physics, P.O. Box 64, University of Helsinki, 00014 Helsinki, Finland\label{aff1}
\and
Helsinki Institute of Physics, Gustaf H{\"a}llstr{\"o}min katu 2, University of Helsinki, 00014 Helsinki, Finland\label{aff2}
\and
Department of Physics and Helsinki Institute of Physics, Gustaf H\"allstr\"omin katu 2, University of Helsinki, 00014 Helsinki, Finland\label{aff3}
\and
INAF-Osservatorio Astronomico di Trieste, Via G. B. Tiepolo 11, 34143 Trieste, Italy\label{aff4}
\and
ESAC/ESA, Camino Bajo del Castillo, s/n., Urb. Villafranca del Castillo, 28692 Villanueva de la Ca\~nada, Madrid, Spain\label{aff5}
\and
INAF-Osservatorio Astronomico di Brera, Via Brera 28, 20122 Milano, Italy\label{aff6}
\and
INAF-Osservatorio di Astrofisica e Scienza dello Spazio di Bologna, Via Piero Gobetti 93/3, 40129 Bologna, Italy\label{aff7}
\and
IFPU, Institute for Fundamental Physics of the Universe, via Beirut 2, 34151 Trieste, Italy\label{aff8}
\and
INFN, Sezione di Trieste, Via Valerio 2, 34127 Trieste TS, Italy\label{aff9}
\and
SISSA, International School for Advanced Studies, Via Bonomea 265, 34136 Trieste TS, Italy\label{aff10}
\and
Dipartimento di Fisica e Astronomia, Universit\`a di Bologna, Via Gobetti 93/2, 40129 Bologna, Italy\label{aff11}
\and
INFN-Sezione di Bologna, Viale Berti Pichat 6/2, 40127 Bologna, Italy\label{aff12}
\and
Dipartimento di Fisica, Universit\`a di Genova, Via Dodecaneso 33, 16146, Genova, Italy\label{aff13}
\and
INFN-Sezione di Genova, Via Dodecaneso 33, 16146, Genova, Italy\label{aff14}
\and
Department of Physics "E. Pancini", University Federico II, Via Cinthia 6, 80126, Napoli, Italy\label{aff15}
\and
INAF-Osservatorio Astronomico di Capodimonte, Via Moiariello 16, 80131 Napoli, Italy\label{aff16}
\and
Dipartimento di Fisica, Universit\`a degli Studi di Torino, Via P. Giuria 1, 10125 Torino, Italy\label{aff17}
\and
INFN-Sezione di Torino, Via P. Giuria 1, 10125 Torino, Italy\label{aff18}
\and
INAF-Osservatorio Astrofisico di Torino, Via Osservatorio 20, 10025 Pino Torinese (TO), Italy\label{aff19}
\and
INAF-IASF Milano, Via Alfonso Corti 12, 20133 Milano, Italy\label{aff20}
\and
INAF-Osservatorio Astronomico di Roma, Via Frascati 33, 00078 Monteporzio Catone, Italy\label{aff21}
\and
INFN-Sezione di Roma, Piazzale Aldo Moro, 2 - c/o Dipartimento di Fisica, Edificio G. Marconi, 00185 Roma, Italy\label{aff22}
\and
Centro de Investigaciones Energ\'eticas, Medioambientales y Tecnol\'ogicas (CIEMAT), Avenida Complutense 40, 28040 Madrid, Spain\label{aff23}
\and
Port d'Informaci\'{o} Cient\'{i}fica, Campus UAB, C. Albareda s/n, 08193 Bellaterra (Barcelona), Spain\label{aff24}
\and
Institute for Theoretical Particle Physics and Cosmology (TTK), RWTH Aachen University, 52056 Aachen, Germany\label{aff25}
\and
Deutsches Zentrum f\"ur Luft- und Raumfahrt e. V. (DLR), Linder H\"ohe, 51147 K\"oln, Germany\label{aff26}
\and
INFN section of Naples, Via Cinthia 6, 80126, Napoli, Italy\label{aff27}
\and
Institute for Astronomy, University of Hawaii, 2680 Woodlawn Drive, Honolulu, HI 96822, USA\label{aff28}
\and
Dipartimento di Fisica e Astronomia "Augusto Righi" - Alma Mater Studiorum Universit\`a di Bologna, Viale Berti Pichat 6/2, 40127 Bologna, Italy\label{aff29}
\and
Instituto de Astrof\'{\i}sica de Canarias, E-38205 La Laguna, Tenerife, Spain\label{aff30}
\and
Institute for Astronomy, University of Edinburgh, Royal Observatory, Blackford Hill, Edinburgh EH9 3HJ, UK\label{aff31}
\and
European Space Agency/ESRIN, Largo Galileo Galilei 1, 00044 Frascati, Roma, Italy\label{aff32}
\and
Universit\'e Claude Bernard Lyon 1, CNRS/IN2P3, IP2I Lyon, UMR 5822, Villeurbanne, F-69100, France\label{aff33}
\and
Institut de Ci\`{e}ncies del Cosmos (ICCUB), Universitat de Barcelona (IEEC-UB), Mart\'{i} i Franqu\`{e}s 1, 08028 Barcelona, Spain\label{aff34}
\and
Instituci\'o Catalana de Recerca i Estudis Avan\c{c}ats (ICREA), Passeig de Llu\'{\i}s Companys 23, 08010 Barcelona, Spain\label{aff35}
\and
Institut de Ciencies de l'Espai (IEEC-CSIC), Campus UAB, Carrer de Can Magrans, s/n Cerdanyola del Vall\'es, 08193 Barcelona, Spain\label{aff36}
\and
UCB Lyon 1, CNRS/IN2P3, IUF, IP2I Lyon, 4 rue Enrico Fermi, 69622 Villeurbanne, France\label{aff37}
\and
Departamento de F\'isica, Faculdade de Ci\^encias, Universidade de Lisboa, Edif\'icio C8, Campo Grande, PT1749-016 Lisboa, Portugal\label{aff38}
\and
Instituto de Astrof\'isica e Ci\^encias do Espa\c{c}o, Faculdade de Ci\^encias, Universidade de Lisboa, Campo Grande, 1749-016 Lisboa, Portugal\label{aff39}
\and
Department of Astronomy, University of Geneva, ch. d'Ecogia 16, 1290 Versoix, Switzerland\label{aff40}
\and
Universit\'e Paris-Saclay, CNRS, Institut d'astrophysique spatiale, 91405, Orsay, France\label{aff41}
\and
INFN-Padova, Via Marzolo 8, 35131 Padova, Italy\label{aff42}
\and
Aix-Marseille Universit\'e, CNRS/IN2P3, CPPM, Marseille, France\label{aff43}
\and
INAF-Istituto di Astrofisica e Planetologia Spaziali, via del Fosso del Cavaliere, 100, 00100 Roma, Italy\label{aff44}
\and
INFN-Bologna, Via Irnerio 46, 40126 Bologna, Italy\label{aff45}
\and
Institut d'Estudis Espacials de Catalunya (IEEC),  Edifici RDIT, Campus UPC, 08860 Castelldefels, Barcelona, Spain\label{aff46}
\and
Institute of Space Sciences (ICE, CSIC), Campus UAB, Carrer de Can Magrans, s/n, 08193 Barcelona, Spain\label{aff47}
\and
School of Physics, HH Wills Physics Laboratory, University of Bristol, Tyndall Avenue, Bristol, BS8 1TL, UK\label{aff48}
\and
University Observatory, LMU Faculty of Physics, Scheinerstr.~1, 81679 Munich, Germany\label{aff49}
\and
Max Planck Institute for Extraterrestrial Physics, Giessenbachstr. 1, 85748 Garching, Germany\label{aff50}
\and
INAF-Osservatorio Astronomico di Padova, Via dell'Osservatorio 5, 35122 Padova, Italy\label{aff51}
\and
Universit\"ats-Sternwarte M\"unchen, Fakult\"at f\"ur Physik, Ludwig-Maximilians-Universit\"at M\"unchen, Scheinerstr.~1, 81679 M\"unchen, Germany\label{aff52}
\and
Institute of Theoretical Astrophysics, University of Oslo, P.O. Box 1029 Blindern, 0315 Oslo, Norway\label{aff53}
\and
Jet Propulsion Laboratory, California Institute of Technology, 4800 Oak Grove Drive, Pasadena, CA, 91109, USA\label{aff54}
\and
Felix Hormuth Engineering, Goethestr. 17, 69181 Leimen, Germany\label{aff55}
\and
Technical University of Denmark, Elektrovej 327, 2800 Kgs. Lyngby, Denmark\label{aff56}
\and
Cosmic Dawn Center (DAWN), Denmark\label{aff57}
\and
Max-Planck-Institut f\"ur Astronomie, K\"onigstuhl 17, 69117 Heidelberg, Germany\label{aff58}
\and
NASA Goddard Space Flight Center, Greenbelt, MD 20771, USA\label{aff59}
\and
Universit\'e de Gen\`eve, D\'epartement de Physique Th\'eorique and Centre for Astroparticle Physics, 24 quai Ernest-Ansermet, CH-1211 Gen\`eve 4, Switzerland\label{aff60}
\and
Laboratoire d'etude de l'Univers et des phenomenes eXtremes, Observatoire de Paris, Universit\'e PSL, Sorbonne Universit\'e, CNRS, 92190 Meudon, France\label{aff61}
\and
SKAO, Jodrell Bank, Lower Withington, Macclesfield SK11 9FT, UK\label{aff62}
\and
Centre de Calcul de l'IN2P3/CNRS, 21 avenue Pierre de Coubertin 69627 Villeurbanne Cedex, France\label{aff63}
\and
University of Applied Sciences and Arts of Northwestern Switzerland, School of Computer Science, 5210 Windisch, Switzerland\label{aff64}
\and
Universit\"at Bonn, Argelander-Institut f\"ur Astronomie, Auf dem H\"ugel 71, 53121 Bonn, Germany\label{aff65}
\and
Aix-Marseille Universit\'e, CNRS, CNES, LAM, Marseille, France\label{aff66}
\and
Dipartimento di Fisica e Astronomia "Augusto Righi" - Alma Mater Studiorum Universit\`a di Bologna, via Piero Gobetti 93/2, 40129 Bologna, Italy\label{aff67}
\and
Department of Physics, Institute for Computational Cosmology, Durham University, South Road, Durham, DH1 3LE, UK\label{aff68}
\and
Universit\'e Paris Cit\'e, CNRS, Astroparticule et Cosmologie, 75013 Paris, France\label{aff69}
\and
CNRS-UCB International Research Laboratory, Centre Pierre Bin\'etruy, IRL2007, CPB-IN2P3, Berkeley, USA\label{aff70}
\and
University of Applied Sciences and Arts of Northwestern Switzerland, School of Engineering, 5210 Windisch, Switzerland\label{aff71}
\and
Institute of Physics, Laboratory of Astrophysics, Ecole Polytechnique F\'ed\'erale de Lausanne (EPFL), Observatoire de Sauverny, 1290 Versoix, Switzerland\label{aff72}
\and
Telespazio UK S.L. for European Space Agency (ESA), Camino bajo del Castillo, s/n, Urbanizacion Villafranca del Castillo, Villanueva de la Ca\~nada, 28692 Madrid, Spain\label{aff73}
\and
Institut de F\'{i}sica d'Altes Energies (IFAE), The Barcelona Institute of Science and Technology, Campus UAB, 08193 Bellaterra (Barcelona), Spain\label{aff74}
\and
European Space Agency/ESTEC, Keplerlaan 1, 2201 AZ Noordwijk, The Netherlands\label{aff75}
\and
DARK, Niels Bohr Institute, University of Copenhagen, Jagtvej 155, 2200 Copenhagen, Denmark\label{aff76}
\and
Universit\'e Paris-Saclay, Universit\'e Paris Cit\'e, CEA, CNRS, AIM, 91191, Gif-sur-Yvette, France\label{aff77}
\and
Space Science Data Center, Italian Space Agency, via del Politecnico snc, 00133 Roma, Italy\label{aff78}
\and
Centre National d'Etudes Spatiales -- Centre spatial de Toulouse, 18 avenue Edouard Belin, 31401 Toulouse Cedex 9, France\label{aff79}
\and
Institute of Space Science, Str. Atomistilor, nr. 409 M\u{a}gurele, Ilfov, 077125, Romania\label{aff80}
\and
Dipartimento di Fisica e Astronomia "G. Galilei", Universit\`a di Padova, Via Marzolo 8, 35131 Padova, Italy\label{aff81}
\and
Instituto de F\'isica Te\'orica UAM-CSIC, Campus de Cantoblanco, 28049 Madrid, Spain\label{aff82}
\and
Institut de Recherche en Astrophysique et Plan\'etologie (IRAP), Universit\'e de Toulouse, CNRS, UPS, CNES, 14 Av. Edouard Belin, 31400 Toulouse, France\label{aff83}
\and
Universit\'e St Joseph; Faculty of Sciences, Beirut, Lebanon\label{aff84}
\and
Departamento de F\'isica, FCFM, Universidad de Chile, Blanco Encalada 2008, Santiago, Chile\label{aff85}
\and
Universit\"at Innsbruck, Institut f\"ur Astro- und Teilchenphysik, Technikerstr. 25/8, 6020 Innsbruck, Austria\label{aff86}
\and
Instituto de Astrof\'isica e Ci\^encias do Espa\c{c}o, Faculdade de Ci\^encias, Universidade de Lisboa, Tapada da Ajuda, 1349-018 Lisboa, Portugal\label{aff87}
\and
Cosmic Dawn Center (DAWN)\label{aff88}
\and
Niels Bohr Institute, University of Copenhagen, Jagtvej 128, 2200 Copenhagen, Denmark\label{aff89}
\and
Universidad Polit\'ecnica de Cartagena, Departamento de Electr\'onica y Tecnolog\'ia de Computadoras,  Plaza del Hospital 1, 30202 Cartagena, Spain\label{aff90}
\and
Kapteyn Astronomical Institute, University of Groningen, PO Box 800, 9700 AV Groningen, The Netherlands\label{aff91}
\and
Caltech/IPAC, 1200 E. California Blvd., Pasadena, CA 91125, USA\label{aff92}
\and
ICSC - Centro Nazionale di Ricerca in High Performance Computing, Big Data e Quantum Computing, Via Magnanelli 2, Bologna, Italy\label{aff93}
\and
Aurora Technology for European Space Agency (ESA), Camino bajo del Castillo, s/n, Urbanizacion Villafranca del Castillo, Villanueva de la Ca\~nada, 28692 Madrid, Spain\label{aff94}
\and
Dipartimento di Fisica - Sezione di Astronomia, Universit\`a di Trieste, Via Tiepolo 11, 34131 Trieste, Italy\label{aff95}
\and
Institut d'Astrophysique de Paris, 98bis Boulevard Arago, 75014, Paris, France\label{aff96}
\and
ICL, Junia, Universit\'e Catholique de Lille, LITL, 59000 Lille, France\label{aff97}}    

  % \date{Received September 15, 1996; accepted March 16, 1997}

% \abstract{}{}{}{}{} 
% 5 {} token are mandatory
 
  \abstract{
   {We study the properties of galaxy cluster 2-point correlation function covariance matrices estimated using the linear-construction (LC) method, which is computationally up to 20 times faster than
   the standard sample-covariance method. Our goal is to assess how well the LC method performs in cosmological parameter estimation compared to the sample covariance.}
   {We use a set of 1000 mock dark matter halo catalogues to compute both the LC-covariance and the sample-covariance estimates in four redshift shells. These numerical matrices are used to fit a theoretical four-parameter model for the covariance. 
   We then use the two fitted covariance models in a likelihood function to estimate two cosmological parameters -- the matter density parameter $\Om$ and the amplitude of the matter density fluctuations $\sigma_8$ -- from the simulated mock catalogues. The purpose of this is to validate the LC-covariance-based model against the sample-covariance model. The catalogues were simulated assuming the spatially flat $\Lambda$CDM cosmology, with $\Om = 0.30711$ and $\sigma_8=0.8288$.
   }
   {We find that the parameter posteriors obtained using the sample- and LC-covariance models agree well with each other and with the simulation cosmology. The two pairs of marginalized constraints are $\Om = 0.307 \pm 0.003$ and $\sigma_8 = 0.826\pm 0.009$ (sample covariance), and $\Om = 0.308 \pm 0.003$ and $\sigma_8 = 0.825 \pm 0.009$ (LC covariance). The posterior widths are the same, and the difference in the median values is less than $0.16\,\sigma$ for both parameters.}
   {}}

   \keywords{large-scale structure of Universe -- cosmological parameters -- Galaxies: statistics -- Galaxies: clusters: general -- Methods: data analysis -- Methods: numerical}

%Gravitational lensing: weak,
   \maketitle
%
%-------------------------------------------------------------------

\nolinenumbers

\section{Introduction}

The ongoing and near-future wide-sky large-scale structure (LSS) surveys, such as \Euclid \citep{EuclidSkyOverview}, the Dark Energy Spectroscopic Instrument experiment \citep{2024AJ....167...62D},
and the \textit{Nancy Grace Roman} Space Telescope
\citep{2019arXiv190205569A},
produce spectroscopic galaxy  catalogues with $10^7$--$10^8$ objects. An important observable is the 2-point correlation function (2PCF) of these objects, which efficiently captures the cosmological information contained in their spatial distribution. However, in order to construct a cosmological likelihood function, we need a covariance matrix of the 2PCF estimate. A straightforward estimate for this would come from averaging over a large number, typically 1000--10\,000, of simulated mock catalogues. This is a computationally challenging task.

For example, the final data release 3 (DR3) of the \Euclid spectroscopic galaxy catalogue will contain $3\times10^{7}$ galaxies. In \Euclid code validation tests, computing a single 2PCF up to the galaxy separations of $r=200\,\hMpc$ from a catalogue of this size with the number density $4\times10^{-3}$\,$h^3$\,Mpc$^{-3}$, using the \Euclid 2PCF code and the standard Landy--Szalay (LS) estimator \citep{1993ApJ...412...64L}, required about one day on a 48 CPU compute node. Thus, calculating 2PCF from 1000 mocks and constructing the covariance matrix estimate would take 1000 days. The method of splitting the random catalogue \citep{2019A&A...631A..73K} reduces this to 100 days, without a significant loss of accuracy or precision. The linear-construction (LC) method \citep[K22 hereafter]{Keihanen22} further reduces the computation time by a factor of 20 (i.e. to 5 days). In a more realistic setup (\citealt{EP-delaTorre}), using the cone geometry and a thin redshift shell, $z=0.9$--$1.1$, and the predicted number density of galaxies in the \Euclid data release 1 (DR1), the brute-force calculation of the covariance matrix from 1000 mocks would take 200 days on a 48 CPU compute node. The split option reduces this to 20 days, and the LC method to less than a day. This remarkable efficiency of the LC method motivates our current study, in which we test, using 1000 mocks and comparing to the standard sample-covariance method, how well the LC method performs in a realistic use case of constraining the cosmological parameters.

This paper is a continuation of the work of K22 in which the linear-construction method was presented. The LC method yields an unbiased numerical estimate for the covariance of a 2PCF estimate computed with the LS estimator. We review the method briefly in Sect.\,\ref{sec:lc-cov}. 
Typically, the covariance matrix is used in cosmological parameter estimation in the form of its inverse, also known as the precision matrix. As is the case with the sample covariance, the direct inverse of the LC covariance is a biased estimate of the precision matrix. In Sect.\,\ref{sec:lc-inverse} we discuss the mathematical properties of the inverse LC covariance and present an improved estimate of the precision matrix.

An alternative to using a numerical covariance matrix estimated from mock measurements is to try to model the covariance analytically, as illustrated by works such as \cite{1994ApJ...426...23F}, \citet{1999ApJ...527....1S}, \cite{1999MNRAS.308.1179M}, \citet{2003ApJ...584..702H}, and \cite{2013PhRvD..87l3504T}. However, purely theoretical covariances may rely on too simplistic approximations to accurately describe realistic galaxy or cluster distributions and survey geometries. A way to improve the situation is to add free parameters to the covariance models and to fit these using measured numerical covariance matrices; see, for example, \citet{2012MNRAS.427.2146X}, \citet{2016MNRAS.462.2681O}, and \citet{2022JCAP...12..022F} for more details. In Sect.\,\ref{sec:modeling} we adopt this approach and introduce our covariance model, the parameters of which can be fitted using the LC covariance. 

In Sect.\,\ref{sec:mocks} we describe the simulated halo catalogues that we used to compute the numerical covariances for the parameter fitting procedure. These two sections rely heavily on the work of \citet{Fumagalli-EP27}, EC24 hereafter, in which the modelling of the covariance of the galaxy cluster 2PCF is studied extensively. Our halo catalogues also rely on the same simulations as those used by EC24.
In Sect.\,\ref{sec:results} we use our covariance models to construct likelihood functions and obtain posterior distributions for two cosmological parameters, and see how the LC covariance compares to the standard sample-covariance method.
Section~\ref{sec:conclusions} contains our conclusions.

In this study we focus on the clustering of galaxy clusters, which, as a test case, offers several benefits as compared to clustering of galaxies: scales are larger, the evolution is mostly linear, and the number of objects is moderate. In addition to being an appealing test case, clustering of clusters is potentially a powerful cosmological probe. See, for example, \citet{marulli:2018}, \citet{marulli:2021}, \citet{moresco:2021}, \citet{lindholm:2021}, \citet{lesci:2022}, \citet{2024A&A...682A.148F}, and \citet{2025A&A...697A..93L} for some recent studies. Even though our tests concern clustering of clusters, nothing prevents, in principle, similar methods from being applied to clustering of galaxies. However, we emphasise that the results of this paper apply only to galaxy clusters, and their applicability to other types of tracers should be verified case-by-case.

\section{The LC Covariance}
\label{sec:lc-cov}
The LC method applies to 2PCF measurements made with the widely used LS estimator \citep{1993ApJ...412...64L}, paired with the split option \citep{2019A&A...631A..73K}. Following the notation of K22, we denote by $\N{d}$ the number of objects in the data catalogue, and parametrise the number of objects in the random catalogue as $\N{r}=M\N{d}$. In the split option the random catalogue is split into $M_{\rm s}$ sub-catalogues, each with $\N{d}M/M_{\rm s}$ objects. Each of the sub-catalogues should be a valid random catalogue (i.e. they should cover the whole survey region, etc.), only with a lower number density. The random pairs needed for the LS 2PCF estimate are then counted only within each sub-catalogue and averaged over the sub-catalogues to reduce computational time. 
\citet{2019A&A...631A..73K} showed that the optimal way to split the random catalogue is such that each sub-random catalogue has the same number of points as the data catalogue, so that $M_{\rm s}=M$. 
This gives the lowest computational cost at a given precision. The split method provides a speedup of a factor of more than ten, with negligible loss of precision. From here on, we assume $M=M_{\rm s}$, and use $M$ to denote both the ratio $\N{r}/\N{d}$ and the split factor.

We write down the split LS estimator in terms of normalised data-data (dd), data-random (dr), and random-random (rr) pair counts:
\begin{align}
    {\rm dd}(\vec{r}) &\coloneqq \frac{{\rm DD}(\vec{r})}{\N{d}(\N{d} - 1)/2} \;, \\
    {\rm dr}(\vec{r}) &\coloneqq \Minv \sum_{i=1}^M \frac{\mathrm{DR}_i(\vec{r})}{N_{\rm d}^2} = \Minv \sum_{i=1}^M \mathrm{dr}_i(\vec{r}) \;, \\
    {\rm rr}(\vec{r}) &\coloneqq \Minv \sum_{i=1}^M \frac{\mathrm{RR}_i(\vec{r})}{N_{\rm d}(N_{\rm d}-1)/2}
  = \Minv \sum_{i=1}^M \mathrm{rr}_i(\vec{r}) \;,
\end{align}
where DD, DR, and RR are the corresponding unnormalised counts, and vector $\Vec{r}$ denotes a separation bin. Using this notation, the split LS estimator is written as
\begin{equation}
    \xi(\vec{r}) = \frac{\mathrm{dd}(\vec{r}) - (2/M)\sum_i \mathrm{dr}_i(\vec{r})}{\Minv \sum_i \mathrm{rr}_i(\vec{r})}+1 \;.
    \label{eq:ls-2pcf}
\end{equation}

The covariance of the 2PCF estimate is by definition
\begin{equation}
    {\rm cov} \left[\xi(\vec{r}_1),\xi(\vec{r}_2)\right] := 
    \left\langle \left[\xi(\vec{r}_1) -\ev{\xi(\vec{r}_1)}\right]
    \left[ \xi(\vec{r}_2) -\ev{\xi(\vec{r}_2)} \right] \right\rangle \;,
    \label{eq:cov-def}
\end{equation}
where the brackets $\left\langle\right\rangle$ denote expectation values. Typically, the covariance of a 2PCF estimate is computed as the sample covariance, using a set of $\N{s}$ mock-catalogue realisations,
\begin{equation}
    {\rm cov} \left[\xi(\vec{r}_1),\xi(\vec{r}_2)\right] \coloneqq
    \frac{1}{\N{s} - 1} \sum_{i=1}^{\N{s}}
    \left[\xi_i(\vec{r}_1)-\bar \xi(\vec{r}_1)\right] \left[\xi_i(\vec{r}_2)-\bar \xi(\vec{r}_2)\right]\;.
    \label{eq:bruteforce}
\end{equation}
Here $\xi_i(\vec{r})$ is a 2PCF estimate measured from the $i$th catalogue realisation, and $\bar \xi(\vec{r})$ is their mean over all the realisations.

By inserting the definition of the LS estimate of Eq.~\eqref{eq:ls-2pcf} into the definition of Eq.~\eqref{eq:cov-def} and by inspecting the covariance of the various pair counts, it can be shown that the covariance is of the form
\begin{equation}
     {\rm cov} \left[\xi(\vec{r}_1),\xi(\vec{r}_2)\right] 
  = \mathbf{A}(\vec{r}_1,\vec{r}_2)+\Minv \mathbf{B}(\vec{r}_1,\vec{r}_2) \;,
  \label{eq:lc}
\end{equation}
where $\mathbf{A}(\vec{r}_1,\vec{r}_2)$ and $\mathbf{B}(\vec{r}_1,\vec{r}_2)$ are independent of $M$, the size of the random catalogue; see K22 for the full derivation. From this result, we notice that if we know the covariance for two values of $M$, we can solve for $\mathbf{A}(\vec{r}_1,\vec{r}_2)$ and $\mathbf{B}(\vec{r}_1,\vec{r}_2)$ and thus construct the covariance for an arbitrary $M$. 

The larger the $M$, the more precise the 2PCF estimate will be. The reference value for the \Euclid spectroscopic survey \citep{EuclidSkyOverview} is $M=50$. K22 show by considering the computational cost of an LS 2PCF estimate as a function of the size of the random catalogue, that by computing a set of 2PCF estimates with $M=1$ and $M=2$ and using Eq.~\eqref{eq:lc} to estimate the sample covariance,  the theoretical computational saving is a factor of $(1 + 3M)/7$. Thus, using the LC method, we save computing time by a factor of roughly 20 compared to computing a set of 2PCF estimates with $M=50$.

The LC covariance is unbiased with respect to the sample covariance, but, as it is computed using fewer random pairs, it has a larger variance. In the tests done in K22, the LC method required 1.2--1.8 times more mock measurements to reach the same precision as the corresponding sample covariance. However, the computational saving for each measurement is so large, that even in this case the use of LC method reduced the total computation time by a factor of 12--18 (not accounting for the production of the mock catalogues), which is still a significant improvement in efficiency. If the LC covariance estimate is too noisy with a given number of mock catalogues, one can choose a larger value $M_a > 1$ for the first set of 2PCF estimates (set $a$). The second set (set $b$) should always be estimated using $M_b = 2M_a$. This way, one can generate two sets of random catalogues of size $M_a$ and compute the random pairs for set $a$ as mean of the two random catalogues and pairs for set $b$ as their union. This procedure reduces the scatter of the LC-covariance estimate, see K22 for details.

The LC method is based on Eq.~\eqref{eq:lc}, which is exact for a 2PCF constructed with a split random catalogue. Similar calculation for a 2PCF obtained with a full random catalogue leads to additional terms of the form $\propto 1/M^2$. It is worth noting that since the accuracy loss introduced by the split option is extremely small (as shown in \citealp{2019A&A...631A..73K}), we can expect the LC covariance to provide a fair estimate for the covariance also in the case where the 2PCF has been computed without splitting the random catalogue.

\section{Inverting the LC covariance}
\label{sec:lc-inverse}
Often in applications, such as cosmological parameter estimation, the quantity needed is the inverse of the covariance matrix, also known as the precision matrix. This poses a problem for the LC-covariance matrix, as it is not positive-definite by construction. To see this, we can write the component matrices of Eq.~\eqref{eq:lc} explicitly,
\begin{align}
  A_{ij} &= 2\hat C^b_{ij} - \hat C^a_{ij} \;, \\
  B_{ij} &= 2M_a\left[\hat C^a_{ij} -\hat C^b_{ij}\right] \;,
\end{align}
where $\hat C^a_{ij}$ and $\hat C^b_{ij}$ are the sample-covariance matrices estimated with $M=M_a$ and $M=M_b$, respectively, and assuming $M_b = 2M_a$. Here and in the following, we use the hat ($\hat C$) to denote the sample covariance. So, the component matrices of the LC covariance are computed as differences between two sample covariances. Depending on how noisy the numerical estimates are and how close to singular the actual covariance matrix is, random fluctuations can cause the smallest eigenvalues to be negative.

Even when the numerical covariance has no negative eigenvalues, its inverse matrix is a biased estimate of the precision matrix; see \citet{Anderson:2003}, \citet{2007A&A...464..399H}, and references therein.
This holds both for the sample covariance of Eq.~\eqref{eq:bruteforce}, and for the LC covariance. The bias can be corrected for in the case of sample covariance by a multiplicative factor\footnote{An alternative to correcting the bias is to account for the uncertainty of the covariance matrix estimate at the likelihood level, see \cite{2016MNRAS.456L.132S} for details.}, known as the Hartlap factor, which depends on the length of the data vector (number of separation bins) $N_{\rm b}$ and on the number of samples available $N_{\rm s}$,
\begin{equation}
    \widehat{\mathbf{C}^{-1}} = \frac{N_{\rm s}-N_{\rm b}-2}{N_{\rm s}-1} \,\mathbf{C}^{-1}\;, \mbox{ for $N_{\rm s} > N_{\rm b} + 2$}\;.
\end{equation}
Here $\mathbf{C}^{-1}$ is the inverse of the sample-covariance matrix. This correction factor, however, does not apply to the LC covariance. The derivation of the Hartlap factor is based on the fact that the sample-precision matrices follow the inverse Wishart distribution. The LC-covariance matrix, however, is a sum of two unrelated sample-covariance matrices, and the corresponding precision matrices do not follow the inverse Wishart distribution.

We can derive an approximate correction for the bias in the LC-precision matrix as follows. Let $\mathbf{C}$ be the true covariance from which the LC covariance deviates by $\mathbf{\Delta}$, $\mathbf{C}_{\rm LC}=\mathbf{C}+\mathbf{\Delta}$. Using the Neumann series expansion for $(\mathbf{C} + \mathbf{\Delta})^{-1} = {\mathbf{C}^{-1}(\mathbf{I + \mathbf{\Delta\, C}^{-1}})}$ the LC-precision matrix becomes
\begin{equation}
\mathbf{C}^{-1}_{\rm LC}
= \mathbf{C}^{-1}
-\mathbf{C}^{-1}\mathbf{\Delta}\;\mathbf{C}^{-1}
+\mathbf{C}^{-1}\mathbf{\Delta}\;\mathbf{C}^{-1}\mathbf{\Delta}\;\mathbf{C}^{-1} \ldots
\end{equation}
Knowing that the LC covariance is unbiased ($\langle\mathbf{\Delta}\rangle=0$), we find the bias-corrected precision matrix as
\begin{equation}
\mathbf{C}^{-1} \approx
\mathbf{C}_{\rm LC}^{-1} -\mathbf{C}^{-1} 
\langle
\mathbf{\Delta}\; \mathbf{C}^{-1} \mathbf{\Delta}
\rangle\,\mathbf{C}^{-1}
\approx
\left(\mathbf{C}_{\rm LC} +\langle\mathbf{\Delta}\; \mathbf{C}^{-1} \mathbf{\Delta}\rangle\right)^{-1}\;.
\label{eq:biascorrection}
\end{equation}
In index notation the bias correction is
\begin{equation}
\langle\mathbf{\Delta}\; \mathbf{C}^{-1} \mathbf{\Delta}\rangle_{il}
= {C}^{-1}_{jk} \langle{\Delta}_{ij}\; {\Delta}_{kl}\rangle
\; .
\end{equation}

A formula for the covariance of the LC covariance was derived in K22,
\begin{align}
   {\rm cov}&\left(C^{\rm LC}_{ij}, C^{\rm LC}_{kl}\right) = \,\langle\Delta_{ij}\Delta_{kl}\rangle \label{eq:covcov_fast2} \nonumber \\
   = & \, \frac{1}{N_{\rm s}-1}\Big[ C_{ik}C_{jl} + C_{il}C_{jk} \nonumber \\
   & + \left(\frac1{2M_a}-\frac1{M}\right) \left(D_{ik}B_{jl}+ B_{ik}D_{jl} +D_{il}B_{jk}+B_{il}D_{jk} \right) \Big] \;,
\end{align}
where $\mathbf{D}=\mathbf{A}+\mathbf{B}/M_a$. With this, the bias correction can be worked into the form
\begin{align}
    \langle\mathbf{\Delta}\; \mathbf{C}^{-1} \mathbf{\Delta} \rangle = & \frac1{N_{\rm s}-1}\Big\{ (1+N_{\rm b})\,\mathbf{C} \label{eq:biascorr1} \nonumber \\
    + & \left(\frac1{2M_a}-\frac1M\right) \big[\mathbf{D}\mathbf{C}^{-1}\mathbf{B} +\mathbf{B}\mathbf{C}^{-1}\mathbf{D} \nonumber \\
    + & \,\mathbf{D}\,{\rm Tr}(\mathbf{C}^{-1}\mathbf{B}) + \mathbf{B}\,{\rm Tr}(\mathbf{C}^{-1}\mathbf{D}) \big]\Big\}
\end{align}
where we have recognised ${\rm Tr}(\mathbf{C}\mathbf{C^{-1}})={\rm Tr}(\mathbf{I})=N_{\rm b}$ as the number of bins.
Note that without the second term, our result reduces to the Hartlap correction in the limit of $N_{\rm s} \gg N_{\rm b} + 2$.

The applicability of the above expression is limited by the fact that it implicitly includes the true precision matrix $\mathbf{C}^{-1}$. Replacing it by $\mathbf{C}_{\rm LC}^{-1}$ is a bad approximation, as this inverse may not even exist. 
One option is finding a stable initial guess and solving Eqs.~\eqref{eq:biascorr1} and ~\eqref{eq:biascorrection} iteratively. For this,
we make rather strong approximations, $\mathbf{C}^{-1}\mathbf{D}\approx \mathbf{I}$ and $\mathbf{B} \approx {\rm constant} \times \mathbf{I}$, and work the bias correction, Eq.~\eqref{eq:biascorrection}, into an approximate, but more practical form
\begin{equation}
\mathbf{C}^{-1} \approx 
\widehat{\mathbf{C}^{-1}_{\rm LC}} \coloneqq
\left\{ \mathbf{C}_{\rm LC} 
+\frac{1+N_{\rm b}}{N_{\rm s}-1}\left[ \mathbf{C}_{\rm LC}+\Big(\frac1{M_a} -\frac2M\Big)\, \mathbf{B}\right]\right\}^{-1}\;. \label{eq:biascorr2}
\end{equation}
Although the expression inside the braces is still not guaranteed to be positive-definite, it is much more stable than $\mathbf{C}_{\rm LC}$ alone. We can now take Eq.~\eqref{eq:biascorr2} as a starting point to retrieve the first estimate for the precision matrix, and then iterate using the more accurate formula of Eqs.~\eqref{eq:biascorr1} and \eqref{eq:biascorrection}. In what follows, we define our convergence criterion such that the element of $\mathbf{C}^{-1}_{\rm LC}$ that changes the most within an iteration step changes by less than 0.1\%. With this criterion we reached convergence within 14--36 steps, depending on the redshift shell in which the LC-covariance matrix was estimated.

In our case of $35\times35$ LC-covariance matrices, the iteration procedure took $\lesssim 3$ ms. The most complex operation within the iteration loop is matrix inversion, which for many methods scales as $\mathcal{O}(n^{2-3})$. This means that computing the LC-precision matrix for a ten times larger data vector would take a few tens of seconds. So, even for significantly larger matrices, the computational cost of this step should be completely negligible, compared to the computation of the matrix itself.

One way to quantify the accuracy of the precision-matrix estimate $\widehat{\mathbf{C}^{-1}}$ is to look at the dot product
\begin{equation}
    \chi^2_i = \left( \vec{\xi}_i -\vec{\bar{\xi}} \right)^T\widehat{\mathbf{C}^{-1}}\left( \vec{\xi}_i -\vec{\bar{\xi}} \right) \;,
    \label{eq:chisq}
\end{equation}
where the vector $\vec{\xi}_i$ is the $i$th 2PCF realisation. This is the form in which the precision matrix enters a Gaussian likelihood function. If $\vec{\xi}_i$ are drawn from a Gaussian distribution and $\widehat{\mathbf{C}^{-1}}$ is an accurate estimate of the precision matrix for the distribution, the values of $\chi^2_i$ (as the name suggests) will follow the $\chi^2$ distribution with $N_{\rm b}$ degrees of freedom. If the same 2PCF realisations are used to compute both the covariance matrix and the $\chi^2$ values, the distribution of these values will be biased. To avoid this, we split our set of 1000 mock 2PCF measurements (see Sect.\,\ref{sec:mocks} for more details) into two sets of 500 realisations. The first set was used to compute the numerical covariance matrices and the second set to compute the $\chi^2$ values. This is also the case for the real 2PCF measurements; the covariance matrices are computed from a set of simulated catalogues that are independent of the real 2PCF measurement.

\begin{figure*}
    \centering
    \includegraphics{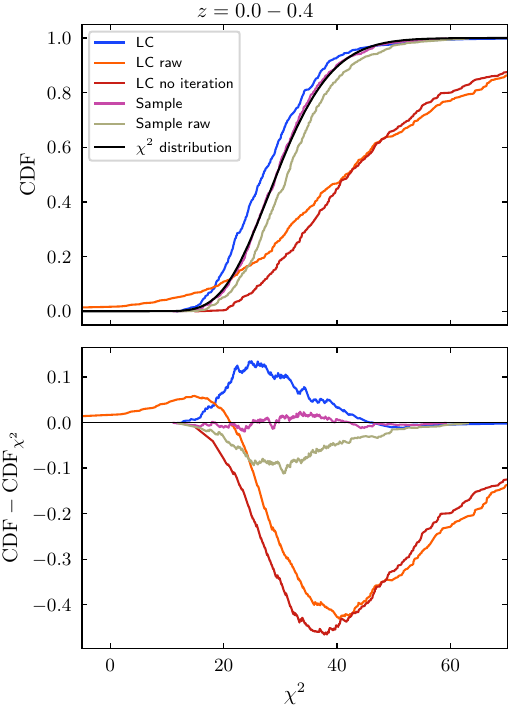}
    \hspace{2mm}
    \includegraphics{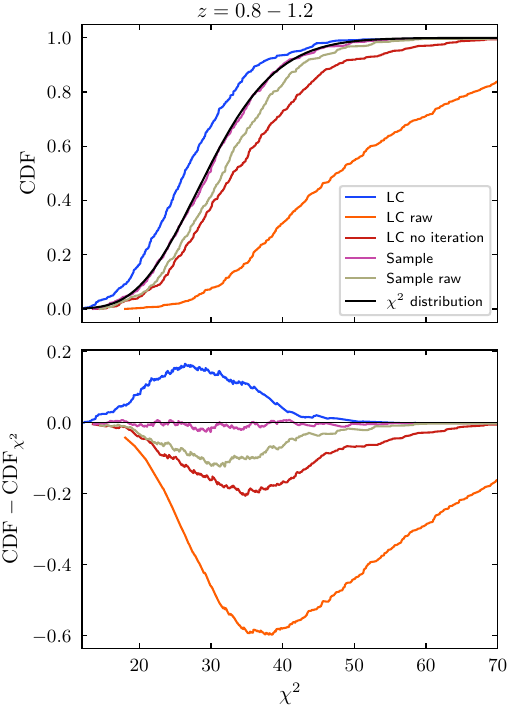}
    \caption{Cumulative distribution of $\chi^2$ values. \emph{Top panels}: distribution of the $\chi^2_i$ values of Eq.~\eqref{eq:chisq} for four flavours of inverse covariance: Hartlap-corrected sample covariance (magenta), the uncorrected LC covariance (orange), the de-biased inverse-LC covariance from Eq.~\eqref{eq:biascorr2} (red), the de-biased inverse-LC covariance after iterating Eq.~\eqref{eq:biascorr1} 14--36 times (blue), and the uncorrected sample covariance (grey). The smooth black line is the theoretical distribution. \emph{Bottom panels}: the difference of each measured CDF with respect to the theoretical prediction. \emph{Left column}: the $z=0.0$--$0.4$ shell. \emph{Right column}: the $z=0.8$--$1.2$ shell.}
    \label{fig:chisq}
\end{figure*}

The cumulative distribution function (CDF) of $\chi^2_i$ values obtained from 500 2PCF realisations using the sample-precision matrix (corrected with the Hartlap factor) and the LC-precision matrix (corrected both using Eq.~\ref{eq:biascorr2} only and also including iteration using Eq.~\ref{eq:biascorr1}) are shown in Fig.~\ref{fig:chisq}, along with the uncorrected LC-inverse matrix. For reference, we also show the sample-precision matrix without the Hartlap correction. The figure shows that the bias corrections indeed improve the LC-precision matrix. Whether the iterative method improves the estimate or not seems to depend on the exact scenario studied (redshift range in our case). However, even after the bias corrections, the LC-precision matrix is still biased compared to the Hartlap-corrected sample-precision matrix or the theoretical expectation. In the case of the $z=0.0$--$0.4$ redshift shell, the figure also demonstrates the effect of a non-positive definite covariance matrix estimate; some of the $\chi^2$ values are negative in the case of the uncorrected inverse LC-covariance matrix.

To verify the bias in the LC-precision matrix, we ran the Kolmogorov-Smirnov (KS) test on the measured $\chi^2$-distributions. We show the $p$-values obtained from the test for both the sample-precision and LC-precision matrices in two redshift shells in Table~\ref{tab:chisq_test}. In both shells, the $p$-values are smaller than $10^{-7}$ for LC-precision matrix, ruling out the hypothesis of the measured $\chi^2$ values following the $\chi^2$ distribution at a high significance. In the same table we also show the difference of the mean and the standard deviation of the $\chi^2$ samples with respect to the theoretical distribution. Whereas the standard deviation of the LC-precision samples does not systematically deviate more than that of sample-precision, the deviation of the mean is significantly larger than the standard error of the mean. The observed bias, even in the corrected precision matrix, suggests that the LC covariance as such is sub-optimal for a likelihood analysis. 

To test the effect of the bias of the LC-precision matrix on cosmological parameter constraints, we performed a likelihood sampling using the iteratively corrected LC-precision matrix and the sample-precision matrix (see Sect.~\ref{sec:likelihood}~and~\ref{sec:results} for description of the exact procedure). In our test scenario, the bias propagated only weakly to the parameter constraints. The differences in the median of the marginalised posteriors for $\Om$ and $\sigma_8$ differed by $0.015\sigma$ and $0.007\sigma$ ($\sigma$ being the width of the 68\% confidence region in the sample-precision case), respectively. Likewise, the width of the marginalised distribution (the 68\% confidence region) for the parameters differed by 2.2\% and 2.9\%, respectively. So, in certain situations, the LC-precision matrix does produce unbiased inference, but its applicability should be checked for each scenario individually.

\begin{table}
\centering
    \caption{Comparison between distribution of our samples of $\chi^2$ values and the theoretical $\chi^2$ distribution, and $p$-values from the Kolmogorov--Smirnov test.}
   \label{tab:chisq_test}
    \begin{tabular}{l c c c c}
        \hline\hline
        \noalign{\vskip 2pt}
        Cov. type & Redshift & $\sqrt{N_{\rm s}}\Delta\mu/\sigma_{\chi^2}$ & $\Delta\sigma/\sigma_{\chi^2}$ & 
        %KS test 
        $p$-value \\
        \noalign{\vskip 2pt}
        \hline
        \noalign{\vskip 1pt}
        Sample & 0.0--0.4 & $-0.084$ & $-0.0047$ & $0.88$ \\
        LC & 0.0--0.4 & $-5.7$ & $0.072$ & $1.3\times 10^{-8}$ \\
        \noalign{\vskip 1pt}
        \hline
        \noalign{\vskip 1pt}
        Sample & 0.8--1.2 & $0.60$ & $0.033$ & $0.87$ \\
        LC & 0.8--1.2 & $-8.3$ & $-0.019$ & $8.0\times 10^{-13}$ \\
        \hline
    \end{tabular}
\end{table}

Even in the case of an unbiased numerical estimate of the precision matrix (such as the Hartlap-corrected sample covariance), it might be computationally too expensive to simulate enough mock catalogues to reach the low enough noise level required for the desired accuracy of the likelihood analysis. Also, from a more theoretical perspective, the covariance matrix is a cosmology-dependent quantity. As shown by EC24, accounting for this cosmology dependence leads to significantly tighter parameter constraints. Thus, searching for ways to model the covariance matrix (and consequently the precision matrix) is well motivated even in the case of unbiased precision matrix estimates.

\section{Modelling}
An alternative to using a numerical covariance matrix as such in a likelihood analysis is to fit a covariance model to the numerical covariance and use the model instead. This avoids all complications related to inverting noisy matrices, discussed in the previous chapter, by involving numerical estimates only at the level of the covariance matrix itself; after the model covariance has been fixed, the precision matrix is simply its inverse matrix, regardless of how the numerical covariance matrices were estimated. Below, we present our 2PCF model and the corresponding covariance model to be used to fit cosmological parameters.

\label{sec:modeling}
\subsection{2PCF}
\label{subsec:2pcf-model}
Our covariance model closely follows the model introduced in EC24, which is a model for the covariance of the real-space 2PCF of dark matter halos. Here we will recap some basic properties of the model. The model is based on the assumption that the halo power spectrum equals the linear matter power spectrum, up to a mass- and redshift-dependent bias. The 2PCF is the Fourier transform of the bias-scaled power spectrum, averaged over a redshift shell and a separation bin. The model assumes a simple conical survey geometry in the redshift shell averaging. The 2PCF model can be written as
\begin{equation}
    \xi_i = \int \frac{\diff k\, k^2}{2\pi^2}\left\langle \bar{b}(z) \sqrt{P_{\rm m}(k, z)}\right\rangle_a^2 W_i(k)\;.
    \label{eq:xi}
\end{equation}
Here the index $i$ labels the separation bin, $\langle \rangle_a$ indicates average over the $a$th redshift shell, $\bar{b}(z)$ is the expected mean bias of the halo sample, $P_{\rm m}(k, z)$ is the linear matter power spectrum, and $W_i(k)$ is the spherical shell window function corresponding to the $i$th separation bin.

To account for the shift and the broadening of the baryon acoustic oscillation (BAO) peak in the halo 2PCF due to the coherent large-scale inflow, we decomposed the matter power spectrum into a smooth (no-wiggle, nw) and an oscillatory (wiggle, w) component and applied a smoothing procedure to the wiggle part of the spectrum.  We first calculated the full linear matter power spectrum using \texttt{CAMB}\footnote{\url{https://camb.info/}} \citep{2000ApJ...538..473L}, and then smoothed the full power spectrum with the Savitzky--Golay filter\footnote{
%%%%%%%%%%%%%%%%
We applied the filter to the logarithm of linear matter power spectrum using an implementation provided by the \texttt{scipy} Python package, \texttt{scipy.signal.savgol\_filter}, with \texttt{poly\_order=1}.
%%%%%%%%%%%%%%%%
} \citep{1964AnaCh..36.1627S} to obtain the no-wiggle spectrum; see \citet{2025JCAP...11..029G} for 13 alternative methods. To get the wiggle spectrum, we simply subtracted the smooth component from the full spectrum. The matter power spectrum was then modelled as the following combination of the two components
\begin{equation}
    P_{\rm m}(k, z) = P_{\rm nw}(k, z) + {\rm e}^{-k^2\Sigma^2(z)}P_{\rm w}(k, z)\;.
\end{equation}
The smoothing process was applied before averaging over the redshift shell, so the no-wiggle spectrum $P_{\rm nw}(k, z)$ and the wiggle spectrum $P_{\rm w}(k, z)$ are redshift dependent. The smoothing parameter $\Sigma^2(z)$ depends on the no-wiggle power spectrum and the BAO scale (the sound horizon scale at the end of the drag epoch), $r_{\rm BAO}$,
\begin{equation}
    \Sigma^2(z) = \int_0^{\,k_{\rm BAO}}\frac{\diff q}{6\pi^2}P_{\rm nw}(q, z)\,[1 - j_0(q\,r_{\rm BAO}) + 2j_2(q\,r_{\rm BAO})]\;.
\end{equation}
Here the integration limit ${k_{\rm BAO}}=2\pi/r_{\rm BAO}$ is the wavenumber corresponding to the BAO scale, and $j_0$ and $j_2$ are the zeroth and the second order spherical Bessel functions, respectively. More details on the theoretical background of the BAO peak processing can be found in EC24 and references therein.

The mean bias was computed as a weighted average over the range of halo masses included in the 2PCF measurement. EC24 considered a covariance model that took into account the covariance of the different mass bins. We, however, only considered a single minimum mass threshold, 
$M=10^{14} h^{-1} \Msun$ (see Sect.~\ref{sec:mocks}),
in which case the bias model reads as
\begin{equation}
    \bar{b}(M, z) = \frac{1}{n(M, z)} \int_M^{\infty} \diff M'\frac{\diff n}{\diff M'}(M', z)\;b(M', z)\;.
\end{equation}
Here $\diff n/\diff M'$ is the halo mass function, $n(M, z)$ is the mean number density of halos above the mass threshold,
\begin{equation}
    n(M, z) = \int_M^{\infty} \diff M'\frac{\diff n}{\diff M'}(M', z)\;,
\end{equation}
and $b(M', z)$ is the expected bias of halos of mass $M'$ at redshift $z$. We used the model of \citet{EP-Castro24} for the halo mass function and \citet{2010ApJ...724..878T} for the halo bias.

Since the goal of our exercise is not to study 2PCF modelling, we minimise any biases in cosmological constraints related to the 2PCF model by forcing the model to agree with the simulated 2PCFs at a fiducial cosmology (the one used to simulate the mock catalogues),
\begin{equation}
    \tilde{\xi_i}(\vec{\Theta}) = \frac{\langle \xi_i \rangle}{\xi_i({\vec{\Theta}_{\rm fid}})}\xi_i({\vec{\Theta}}) \;.
    \label{eq:xi-rescaled}
\end{equation}
Here $\xi_i(\vec{\Theta})$ is the ``raw'' 2PCF model of Eq.~\eqref{eq:xi} and $\langle \xi_i \rangle$ is the mean of the measured 2PCFs in the $a$th redshift shell. Vector $\vec{\Theta}$ denotes parameters of the cosmological model, and the subscript ``fid'' refers to their values at the simulation cosmology. This rescaling ensures a perfect fit at the fiducial cosmology but retains the correct cosmology dependence in the 2PCF model.

\subsection{Covariance}
\label{subsec:cov-model}
A covariance model based on the 2PCF model of Eq.~\eqref{eq:xi} can be obtained by Fourier transforming the corresponding halo power spectrum covariance. Following this procedure, we arrive at the following covariance model
\begin{align}
\begin{split}
    C_{aij} = & \frac{2}{V_a}\int \frac{\diff k \,k^2}{2\pi^2}\left[ \left\langle \bar{b}^2\Pm\right\rangle_a + \left\langle \frac{1}{\bar{n}} \right\rangle_a  \right]^2 W_i(k) W_j(k) \\
    & + \frac{2}{V_aV_i}\,\int \frac{\diff k\, k^2}{2\pi^2} \left\langle \bar{b}^2\Pm\right\rangle_a \left\langle \frac{1}{\bar{n}} \right\rangle_a^2 W_i(k)\,\delta_{ij}\;.
    \label{eq:cov-model}
\end{split}
\end{align}
Here $V_a$ is the volume of the redshift shell, $V_i$ is the volume of the spherical shell corresponding to the $i$th separation bin, and $\langle 1/\bar{n} \rangle_a$ is the shot noise averaged over the redshift shell. The first line represents the Gaussian covariance, and the second line the lowest-order non-Gaussian contribution. The model ignores the higher-order (i.e. three- and four-point) correlation functions as well as correlations between the redshift shells and the super-sample covariance. See EC24 and references therein for more discussion on these aspects.

In EC24, the model of Eq.~\eqref{eq:cov-model} was shown to underestimate the numerical covariance by $\sim$\,30\% on the main diagonal and by $\sim$\,50\% on the off-diagonals above redshifts of 0.4. Our results are similar, see Sect.~\ref{sec:results}. Our case of the main diagonal is slightly more ambiguous; the model underestimates numerical covariance in some redshift/scale combinations and overestimates in others at the level of $\sim$\,10--20\%. At redshifts above 0.4, the off-diagonals are systematically underestimated at the level of $\sim$\,40--60\%.

Motivated by these shortcomings, EC24 extended the covariance model by introducing three free parameters to the model. Bias $\bar{b}$ was replaced by $\beta\bar{b}$, the first shot noise term $\langle 1/\bar{n}\rangle_a$ by $\langle (1 + \alpha)/\bar{n}\rangle_a$, and the second shot noise term $\langle 1/\bar{n}\rangle_a^2$ by $\langle (1 + \gamma)/\bar{n}\rangle_a^2$. These parameters were fitted to the measured numerical covariance within each redshift shell. This significantly improved the agreement between the model and the numerical covariance.
\begin{figure}
    \centering
    \includegraphics{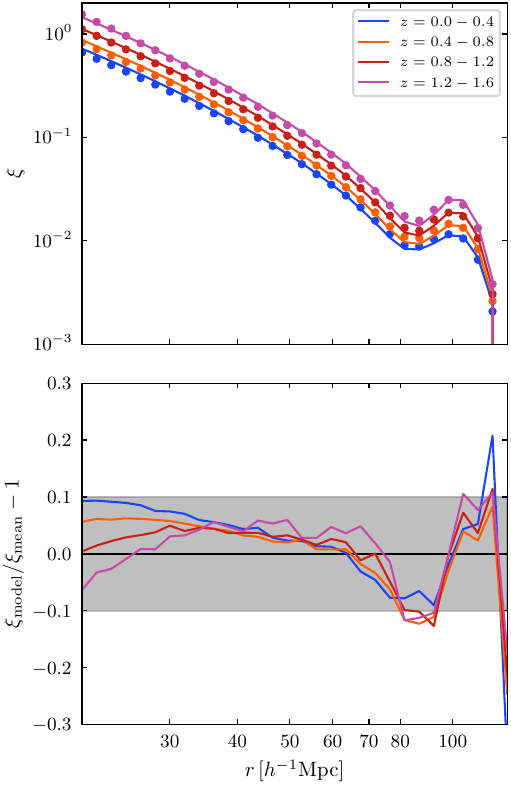}
    \caption{Comparison of the modelled 2PCF, $\xi(r)$, and the mean measured from 1000 simulations. \emph{Top panel}: the mean measured (points) and model (solid lines) 2PCF in four redshift shells. \emph{Bottom panel}: the relative difference between the model and the measurement. The grey box shows the 10\% region.}
    \label{fig:xi}
\end{figure}

Our goal is to see how the LC covariance compares to the sample covariance when it is used to fit parameters of the covariance model. The fitting procedure of EC24, however, only applies to the sample covariance. For details and the full derivation of the method, see \citet{2022JCAP...12..022F}. The LC covariance is a linear combination of two sample covariances (at $M=1$ and $M=2$), which prevents the method from being applied to it, and thus does not allow our desired comparison. Because of this, we parametrise our model in a slightly different way. If we expand the square of the binomial in the first term of Eq.~\eqref{eq:cov-model}, we see that the covariance model is a sum of four terms. We give each term a free coefficient and treat the covariance matrix as a linear combination of four component matrices,
\begin{align}
    C_{a\,ij} = & \ \ p_{a\,1} \ \frac{2}{V_a}\int \frac{\diff k \,k^2}{2\pi^2}\left\langle \bar{b}^2\Pm\right\rangle_a^2 W_i(k) W_j(k) \label{eq:cov-model2} \nonumber \\
    & + p_{a\,2} \ \frac{2}{V_a}\int \frac{\diff k \,k^2}{2\pi^2}2\left\langle \bar{b}^2\Pm\right\rangle_a \left\langle \frac{1}{\bar{n}} \right\rangle_a  W_i(k) W_j(k) \nonumber \\
    & + p_{a\,3} \ \frac{2}{V_a}\int \frac{\diff k \,k^2}{2\pi^2}\left\langle \frac{1}{\bar{n}} \right\rangle_a ^2 W_i(k) W_j(k) \nonumber \\
    & + p_{a\,4} \ \frac{2}{V_aV_i}\, \int \frac{\diff k\, k^2}{2\pi^2} \left\langle \bar{b}^2\Pm\right\rangle_a \left\langle \frac{1}{\bar{n}} \right\rangle_a^2 W_i(k)\,\delta_{ij} \;,
\end{align}
or writing more compactly, for each redshift shell $a$,
\begin{equation}
\mathbf{C} = \sum_{k=1}^4 p_{k}\,\mathbf{C}_{k} \;.
\label{eq:cov-model3}
\end{equation}

We performed a standard weighted least squares fit to the numerical covariance to determine the parameters $p_k$. We used as weights the inverse of the estimated standard deviation of each element of the numerical covariance matrix. These standard deviations can be read directly from Eq.~\eqref{eq:covcov_fast2}. Following EC24, the parameters $p_k$ were fitted only at the fiducial cosmology so that the covariance model retained the correct cosmology dependency. 
\begin{figure*}
    \centering
    \includegraphics{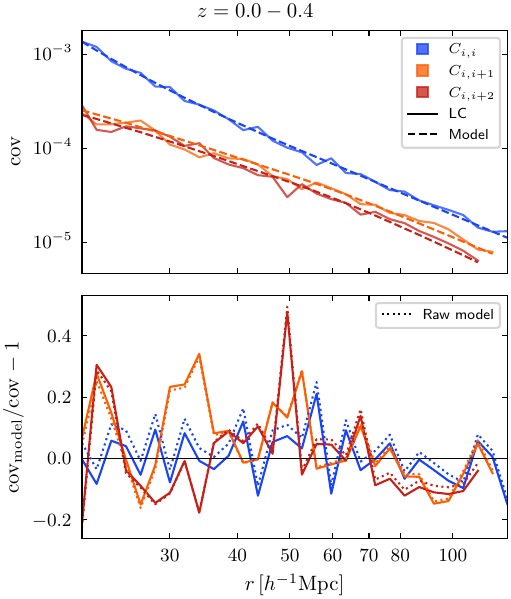}
    \hspace{2mm}
    \includegraphics{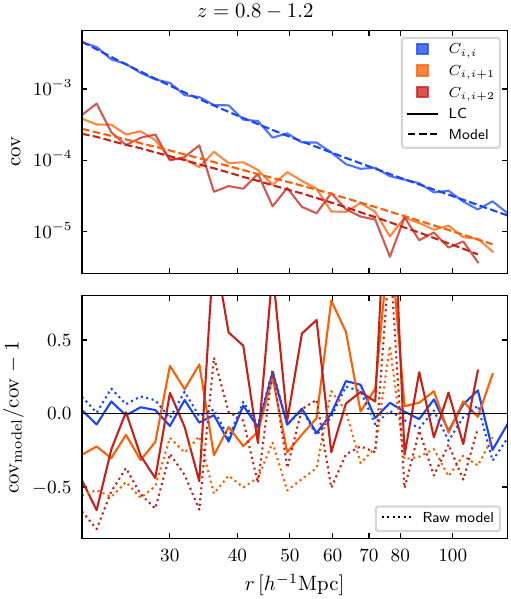}
    \caption{Comparison of the covariance model and the numerical LC covariance. \emph{Top panels}: the first three diagonals of the LC covariance (solid lines), and the model covariance (dashed lines) from Eq.~\eqref{eq:cov-model2} with parameters $p_k$ for LC given in Table~\ref{tab:params-cov}. \emph{Bottom panels}: the relative difference of the model with respect to the LC covariance (solid lines), in addition to the difference of the ``raw'' covariance model (the free parameters $p_k$ set to unity in Eq.~\ref{eq:cov-model2} or Eq.~\ref{eq:cov-model3}) with respect to the numerical LC covariance (dotted lines). \emph{Left column}: the $z=0.0$--$0.4$ shell. \emph{Right column}: the $z=0.8$--$1.2$ shell.}
    \label{fig:cov-diags-lc}
\end{figure*}
\begin{figure*}
    \centering
    \includegraphics{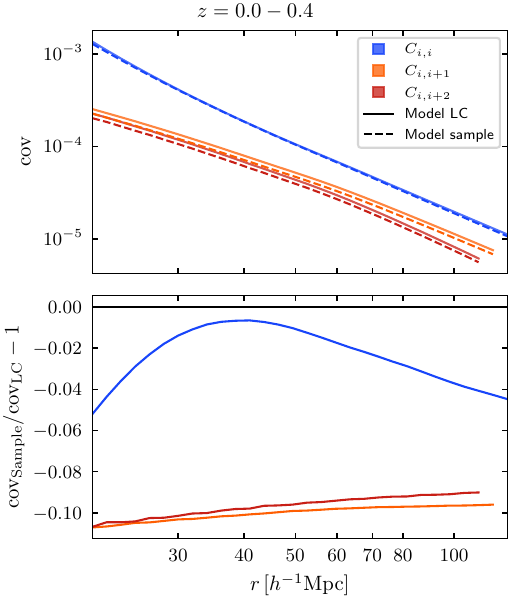}
    \hspace{2mm}
    \includegraphics{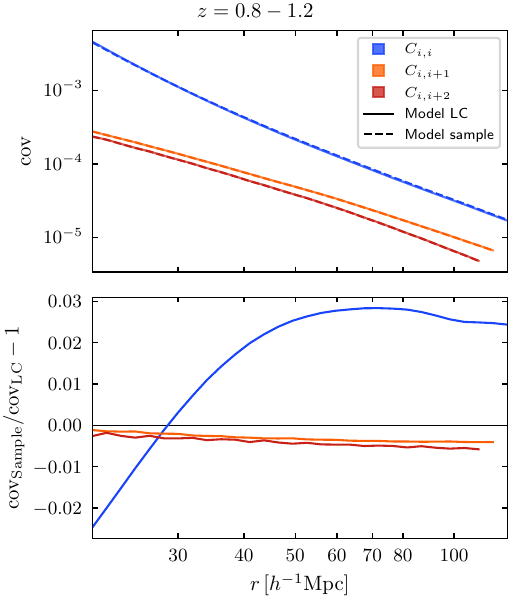}
    \caption{Comparison of covariance models fitted to either the LC covariance or the sample covariance. \emph{Top panels}: the first three diagonals of the model corresponding to the LC covariance (solid lines) and the model corresponding to the sample covariance (dashed lines). \emph{Bottom panels}: the relative difference of the LC-covariance model with respect to the sample-covariance model. \emph{Left column}: the $z=0.0$--$0.4$ shell. \emph{Right column}: the $z=0.8$--$1.2$ shell.}
    \label{fig:cov-diags-diff}
\end{figure*}

\subsection{Likelihood}
\label{sec:likelihood}
To study the usefulness of the LC method in a cosmological analysis, we compared cosmological parameter posteriors obtained by using in the likelihood function either the sample covariance or the LC covariance. 
%This means taking the covariance-model parameters $p_k$ from Table~\ref{tab:params-cov}.
For each case, we fitted the covariance-model parameters $p_k$ to the corresponding numerical covariance and used the fitted parameters to construct the model covariance entering the likelihood.
In this analysis we assumed the likelihood to be Gaussian in the 2PCF,

\begin{align}
    \begin{split}
        \ln{\mathcal{L}\left(\datab|\modelb, \covb\right)} = &-\frac{1}{2}\left\{ \left[ \datab -\modelb \right]^{\rm T} \invcovb \left[ \datab -\modelb \right]  \right\}\, \\
        &- \frac{1}{2} \ln{\left[ \left( 2\pi \right)^{N_{\rm b}} \det \covb \right]}  \;. 
    \end{split}
\end{align}
Here vector $\thetab$ denotes the sampled parameters and, for each redshift shell, $\datab$ is the data vector, vector $\modelb$ is the 2PCF model, $\covb$ is the covariance matrix model (defined by the values of $p_k$), and $N_{\rm b}$ is the length of the data vector (i.e. the number of separation bins). Both the 2PCF model and the covariance model depend on the cosmological parameters. 

Each individual 2PCF measurement is subject to the effects of cosmic variance. To suppress these effects, we computed the final likelihood as the average over all the mock measurements,
\begin{equation}
    \ln{\mathcal{L}} = \frac{1}{N_{\rm s}} \sum_{i=1}^{N_{\rm s}} \ln{\mathcal{L}_i} \;.
    \label{eq:like-mean}
\end{equation}
Here $\mathcal{L}_i$ is the likelihood computed from the $i$th mock measurement and $N_{\rm s}$ is the number of such measurements.

%--------------------------------------------------------------------
\section{Mock 2PCF measurements}
\label{sec:mocks}
To compute our numerical covariances, both sample and LC, we need a set of mock catalogues. In our analysis, we used the same set of simulated dark matter halo catalogues as EC24. This is a set of 1000 past light cones produced with the \texttt{PINOCCHIO}\footnote{\url{https://github.com/pigimonaco/Pinocchio}} (PINpointing Orbit-Crossing Collapsed HIerarchical Objects) algorithm \citep{2002MNRAS.331..587M,2017MNRAS.465.4658M}. The light cones cover a spherical cap with an area of 10\,313 ${\rm deg}^2$ and a redshift range $z = 0$--$2.5$. The minimum halo mass in the catalogues is $3.16  \times 10^{13}h^{-1} \Msun$, which corresponds to the halo being sampled with more than 50 particles. However, following EC24, in our analysis we only included halos with a virial mass larger than $10^{14} h^{-1} \Msun$ in the redshift range $z=0$--$1.6$. With this selection, the final catalogues contain $\sim 1.4\times 10^5$ halos each. The simulations assume a spatially flat $\Lambda$CDM cosmology with parameter values from \citet{2014A&A...571A..16P}, as listed in Table~\ref{tab:cosmo_pinocchio}.
\begin{table}
\centering
    \caption{Parameters of the flat $\Lambda$CDM cosmology used to simulate the mock catalogues.}
   \label{tab:cosmo_pinocchio}
    \begin{tabular}{c c c c c}
        \hline\hline
        \noalign{\vskip 2pt}
         $\Om$ & $\Omega_{\rm b}$ & $h$ & $n_{\rm s}$ & $\sigma_8$ \\
         \noalign{\vskip 2pt}
         \hline
         \noalign{\vskip 1pt}
         0.30711 & 0.048254 & 0.6777 & 0.96 & 0.8288 \\
         \hline
    \end{tabular}
\end{table}
\begin{table*}
\centering
    \caption{Values of the parameters $p_{k}$ of the covariance model fitted to both the sample-covariance and the LC-covariance matrix.}
    \label{tab:params-cov}
    \begin{tabular}{l@{\hskip 1.5em} c@{\hskip 2em} r@{\hskip 1.5em} c@{\hskip 1.5em} c@{\hskip 1.5em} c}
        \hline
        \hline
         \noalign{\vskip 2pt}
        Cov.\ type & Redshift & \multicolumn{1}{c}{$p_{1}$} & $p_{2}$ & $p_{3}$ & $p_{4}$  \\
         \noalign{\vskip 2pt}
        \hline
        \noalign{\vskip 1pt}
        Sample &  0.0--0.4 & $0.84 \pm 0.03$ & $0.97 \pm 0.05$ & $1.04 \pm 0.03$ & $0.70 \pm 0.09$ \\
        LC & 0.0--0.4 & $0.88 \pm 0.07$ & $1.13 \pm 0.10$ & $0.92 \pm 0.07$ & $0.95 \pm 0.18$ \\
         \noalign{\vskip 1pt}
        \hline
         \noalign{\vskip 1pt}
        Sample & 0.4--0.8 & $0.82 \pm 0.06$ & $1.40 \pm 0.06$ & $0.90 \pm 0.02$ & $1.02 \pm 0.07$ \\
        LC & 0.4--0.8 & $0.72 \pm 0.14$ & $1.45 \pm 0.14$ & $0.91 \pm 0.06$ & $0.98 \pm 0.13$ \\
        \noalign{\vskip 1pt}
        \hline
        \noalign{\vskip 1pt}
        Sample & 0.8--1.2 & $0.46 \pm 0.23$ & $1.89 \pm 0.10$ & $0.92 \pm 0.02$ & $0.82 \pm 0.05$ \\
        LC & 0.8--1.2 & $0.49 \pm 0.56$ & $1.89 \pm 0.25$ & $0.88 \pm 0.05$ & $0.90 \pm 0.09$ \\
        \noalign{\vskip 1pt}
        \hline
        \noalign{\vskip 1pt}
        Sample & 1.2--1.6 & $-2.27 \pm 1.46$ & $2.22 \pm 0.23$ & $0.89 \pm 0.02$ & $0.75 \pm 0.03$ \\
        LC & 1.2--1.6 & $-0.87 \pm 3.94$ & $2.43 \pm 0.61$ & $0.86 \pm 0.04$ & $0.75 \pm 0.07$ \\
        \hline
    \end{tabular}
\end{table*}

To avoid issues with the modelling of the halo mass functions, the halo masses have been scaled to match a known mass function; see \citet{Fumagalli21} for details of this rescaling. The only difference between our catalogues and those used in EC24 is the target mass function for the rescaling. In EC24, it was that of \citet{2016MNRAS.456.2486D}, whereas our catalogues follow the mass function of \citet{EP-Castro24}. \citet{2025arXiv251013509F} give a more detailed description of our catalogues.

We performed the real-space 2PCF measurements in 35 logarithmically spaced bins over the scales of $r=20$--$130\,\hinvmpc$. For each data catalogue we produced a random catalogue by sampling right ascension and declination to cover the spherical cap and redshift to match the mean tabulated redshift distribution of the data catalogues. The random catalogues for the sample covariance were 50  times the size of the data catalogues to be in line with the \Euclid reference value of $M=50$. The corresponding random catalogues for the LC covariance had $M=1$ and $M=2$.

All of our 2PCF measurements were performed using the \Euclid 2PCF estimator code described in \citet{EP-delaTorre}. We measured the 2PCFs in four consecutive redshift shells, covering $z=0.0$--$1.6$ with $\Delta z = 0.4$. 
EC24 considered one additional shell, $z=1.6$--$2.0$, but this shell turned out to be so sparsely populated that the $M=1$ covariance matrix was too noisy to be useful (the first two bins at the smallest scales contained no $RR$ pairs in the $M=1$ case for some catalogue realisations). This problem could have been avoided by using larger values for $M$ in this shell ($M=2$ and $M=4$, for example) but we did not want to complicate the analysis by involving LC covariances of varying precision levels.

\section{Results}
\label{sec:results}
Our measured 2PCF averaged over the 1000 realisations and the corresponding model are shown for each redshift shell in Fig.~\ref{fig:xi}. The model 2PCF have not been rescaled according to Eq.~\eqref{eq:xi-rescaled}; instead, they show the agreement between the purely theoretical model of Eq.~\eqref{eq:xi} and the measured 2PCF. Apart from few bins close to the BAO peak, the agreement is within the 10\% range in all the redshift shells.

The fitted values for the parameters $p_k$ in each redshift shell are given in Table~\ref{tab:params-cov}. The error estimates correspond to the estimated variance of the elements of the numerical covariance matrices propagated through the standard least squares fit error estimation. 
The main diagonal and the first two off-diagonals of the actual covariance model produced by these parameter values are shown in Fig.~\ref{fig:cov-diags-lc}, along with the numerical LC covariance measured from our mock catalogues and used to fit the parameters. We show the comparison in two redshift shells: $z=0.0$--0.4 and $z=0.8$--1.2. In the difference plot, we also show a comparison of the LC covariance with the ``raw'' covariance model (the free parameters $p_k$ set to unity) to visualise the improvements of the extended model, especially for off-diagonals at higher redshifts. There are some residuals left but the situation is much better than in the case of the ``raw'' model. The results are in line with those of EC24 for the sample covariance and indicate the higher noise level in the LC covariance.

Figure~\ref{fig:cov-diags-diff} shows a comparison of the covariance model of Eq.~\eqref{eq:cov-model3} fitted to either the numerical LC covariance or the sample covariance, in the same redshift shells as in Fig.~\ref{fig:cov-diags-lc}. These two models correspond to the parameter values, $p_k$, in Table~\ref{tab:params-cov} and are the ones used in our likelihood analysis. The models agree well, differences being of the order of a few to ten per cent, depending on the diagonal, the redshift shell and scale. This demonstrates the suitability of the LC covariance for such a fitting procedure.

\begin{figure}[t]
    \centering
    \includegraphics{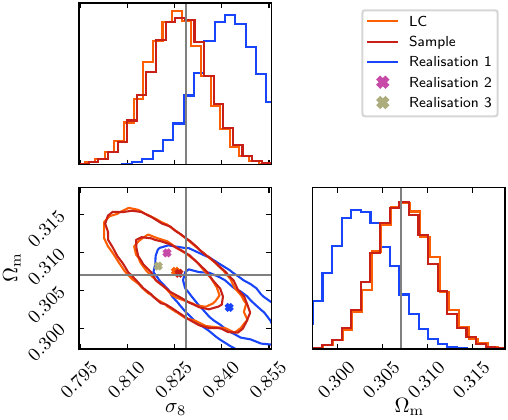}
    \caption{Posterior distribution for $\Om$ and $\sigma_8$ obtained from four redshift shells. The LC covariance is shown in orange, and the sample covariance in red. The grey crosshair shows the values used in the simulation. The blue contours show the posterior in the case of the sample covariance, when constructing a likelihood from a single 2PCF realisation instead of the mean over the 1000 light cones. Crosses of corresponding colours show the median of each distribution. We have also plotted medians from two additional individual realisations to further illustrate their spread. The 2D contours correspond to the 68\% and 95\% confidence regions.}
    \label{fig:posterior}
\end{figure}

For the cosmological parameter estimation, we performed the likelihood sampling using the \texttt{Nautilus}\footnote{\url{https://github.com/johannesulf/nautilus}} sampler \citep{2023MNRAS.525.3181L} within the \texttt{CosmoSIS}\footnote{\url{https://github.com/joezuntz/cosmosis}} framework \citep{2015A&C....12...45Z}. Figure~\ref{fig:posterior} shows the posterior distributions we obtained for $\Om$ and $\sigma_8$ when using covariance models fitted to the sample covariance and the LC covariance. We combined all four redshift shells into a single likelihood simply by taking the product of the individual likelihoods. This is justified by EC24, where it was shown that the covariance between different redshift shells is negligible. We also show a posterior obtained by using a single 2PCF realisation, without averaging the likelihood over all the realisations (as defined in Eq.~\ref{eq:like-mean}), along with median values from two additional realisations. This highlights how small the shift caused by the LC covariance, with respect to the case of the sample covariance, is compared to the variance among different realisations. From the averaged likelihoods we obtain the marginalised constraints of $\Om = 0.307 \pm 0.003$, $\sigma_8 = 0.826\pm 0.009$ (sample covariance)
 and $\Om = 0.308 \pm 0.003$, $\sigma_8 = 0.825 \pm 0.009$ (LC). The difference in the median value is $0.12\,\sigma$ for $\Om$ and $0.16\,\sigma$ for $\sigma_8$. 
 
 These posteriors were obtained using covariance models that depended on the sampled cosmological parameters. We also ran samplings with fixed covariance models (evaluated at the fiducial cosmology) and verified the EC24 result that in this case the posteriors are significantly wider. The additional information contained in the cosmology-dependent covariance model is attributed to the shot-noise term, which depends on the expected cluster number density (and thus on the cluster mass function and consequently on cosmological parameters). This term is not part of the 2PCF model.
 For a more detailed discussion of this effect, see \cite{2025arXiv251013509F}.

 In our tests we used the same set of 2PCF realisations to compute the LC- and the sample-covariance matrices. As noted earlier, the LC covariance has a larger variance than the sample covariance if computed using the same number of 2PCF realisations. In this scenario, the computational saving of the LC method is roughly a factor of 20. Had we used more realisations to compute the LC covariance, the difference the two matrices make in a cosmological analysis would have been even smaller than what we observed. As noted earlier, requiring the same noise level as with the corresponding sample-covariance estimate would result in a speed-up of a factor of 12--18 with the LC method. However, since the computational cost of the extra mock catalogues might not be negligible, their cost and the specific precision requirements should be judged against the cost of estimating the 2PCFs when deciding the covariance estimation method.

\section{Conclusions}
\label{sec:conclusions}
In this paper, we continued the work of K22, in which the linear-construction method for computing the 2PCF covariance matrix was introduced. In the present work we studied the suitability of the LC covariance for cosmological parameter estimation. 

A well-known fact is that the inverse of the sample-covariance-matrix estimate is a biased estimate of the inverse-covariance matrix, also known as the precision matrix. This bias can be corrected for by a multiplicative factor. However, this factor does not apply to the LC-covariance matrix. To mitigate this, we derived an approximate correction factor specific to the LC method. By comparing the distribution of $\chi^2$ values obtained by using the LC-precision matrix to the theoretical $\chi^2$ distribution, using the Kolmogorov-Smirnov test, we find that while the correction does indeed improve the precision-matrix estimate, the result is still biased. This suggests that the numerical LC covariance as such is sub-optimal as a part of a likelihood function. However, even the bias-corrected sample-covariance-based numerical precision matrix estimate might be problematic due to the numerical noise.

As an alternative to the direct application in the likelihood function, we studied using the LC covariance to fit parameters of a dark matter halo 2PCF covariance model. Our goal was to replicate the analysis of EC24, but using the LC covariance, and to compare the results to the sample covariance. We had to make a small modification in the formulation of the free parameters of the covariance model because the fitting procedure of EC24 does not apply to the LC covariance. Using this modified parametrisation, we find the models fitted to both the sample covariance and the LC covariance to recover the numerical (sample covariance) at a level of a few tens of per cent, depending on which redshift bin and main or off-diagonal is considered. The results are similar to those obtained in EC24, with the LC-covariance-fitted model having slightly larger residuals.

Finally, we studied how the differences in the two covariance models affected cosmological parameter estimation. We estimated the posterior distributions of the cosmological parameters $\Om$ and $\sigma_8$ using both models in our likelihood function. We find the shift in the 2D posteriors to be completely negligible compared to the variance among different light-cone realisations. We obtain the marginalized constraints of $\Om = 0.307 \pm 0.003$ and $\sigma_8 = 0.826\pm 0.009$ (sample covariance), and $\Om = 0.308 \pm 0.003$ and $\sigma_8 = 0.825 \pm 0.009$ (LC covariance). The posterior widths are the same, and the difference in the median values is less than $0.16\,\sigma$ for both parameters. These posteriors were computed as a mean likelihood over 1000 2PCF realisations. We verified by plotting parameter medians from three individual 2PCF realisations that the shifts in the median of the distributions far exceeded the one introduced by the LC covariance. So, in the scenario studied in this paper, the LC-covariance method produced essentially the same parameter constraints as the sample covariance, but at a 20 times lower computational cost.

In this paper we studied the clustering of dark matter halos, which is the basis for the clustering of galaxy clusters. In principle, nothing prevents applying the LC covariance for clustering of, for example, galaxies. The only requirements for a similar analysis are (i) a model for the 2PCF covariance and (ii) a set of simulations for measuring the numerical covariance to be used to fit any free parameters in the covariance model. However, galaxies are affected more, for example, by non-linear growth of structure and by complex baryonic physics, making the covariance modelling significantly more difficult. On the other hand, galaxies also have a significantly higher number density than clusters, which would make the computational savings of the LC method even more appealing. The results of this paper cannot, a priori, be extrapolated to tracers other than galaxy clusters, but the suitability of the LC method to galaxy samples would be an interesting topic for future research.

\begin{acknowledgements}
    This work has been supported by the Research Council of Finland grant 347088, and the Jenny and Antti Wihuri Foundation. The authors wish to acknowledge CSC – IT Center for Science, Finland, for computational resources.
    \AckEC 
\end{acknowledgements}

%\jv{Referensseissä tarvitaan viimeiseksi riviksi aina validi adsurl = \{https://ui.adsabs.harvard.edu/abs/XXXXXXXXXX\}-rivi!}

\bibliographystyle{aa_url}
\bibliography{Euclid.bib,bibliography.bib}

\label{LastPage}

\end{document}